\documentclass[12pt]{iopart}
\expandafter\let\csname equation*\endcsname\relax
\expandafter\let\csname endequation*\endcsname\relax
\usepackage{graphicx,animate}
\usepackage{algorithm}
\usepackage{algpseudocode}
\usepackage{latexsym}
\usepackage{amssymb}
\usepackage{amsmath}
\usepackage{xfrac}

\begin{document}

\title[Energy-Flexible Network for Dual-Energy CT Image Reconstruction]{Energy-Flexible Network (EF-Net) for Sparse-View Dual-Energy CT Image Reconstruction}

\author{Donghyeon Lee$^1$\footnote{Present address: The Russell H. Morgan Department of Radiology and Radiological Science, Johns Hopkins University School of Medicine, Baltimore, Maryland, U.S.A.}, Uijin Jeong$^1$, Sungho Yun$^1$, Joonil Hwang$^1$, and Seungryong Cho$^{1,\star}$}

\address{Department of Nuclear and Quantum Engineering, Korea Advanced Institute of Science and Technology, Daejeon, 34141, Korea}
\ead{scho@kaist.ac.kr}
\vspace{10pt}
\begin{indented}
\item[]March 2023
\end{indented}

\begin{abstract}
In dual-energy computed tomography (DECT), the X-ray tube energy pair often changes depending on the target organ or patient obesity. In practice, it makes difficult to apply deep learning (DL) based algorithms for image reconstruction since most of the existing DL-based algorithms are trained to be used for dedicated X-ray tube energies. In this paper, we propose 1) an energy flexibility training (EFT) method, which makes a network applicable for data measured at various X-ray tube energies between two trained energies, and 2) an effective dual-domain convolutional neural network for image reconstruction. The proposed network is derived from the regularized version of the primal-dual hybrid gradient algorithm, so its architecture has an unfolded iterative dual-domain structure. For validation, we generated datasets from a lab-made polychromatic X-ray simulator. The proposed method showed promising results not only at the trained X-ray tube energies but also at untrained X-ray tube energies outperforming an iterative algorithm and other DL-based algorithms.
\end{abstract}

%\maketitle

\section{Introduction}

Dual-energy computed tomography (DECT) is a widely used imaging modality in various clinical practices since it has the potential to improve the diagnostic performance of CT by providing additional material-selective information \cite{agostini2019dual, grajo2016dual}. Applications of dual-energy CT images are diverse depending on the diagnostic purposes: virtual non-contrast imaging \cite{heo2016effectiveness, yamada2014radiotherapy}, bone removal in vascular imaging \cite{schulz2012automatic}, characterization of renal stones \cite{lestra2016applications, primak2007noninvasive}, differentiation of hemorrhage in neuroimaging \cite{gupta2010evaluation}, etc. In recent years, deep learning (DL) approaches have been investigated extensively in the CT imaging field due to their remarkable performance \cite{adler2018learned, chen2018learn, jin2017deep, zhang2020metainv}. Particularly under scanning conditions such as sparse-view scan or low tube-current scan, DL approaches have successfully shown much improved image quality. However, they generally require massive data for training, which is quite challenging in DECT since a pair of X-ray tube energies for low-energy and high-energy can be changed according to the target organ or patient obesity \cite{doerner2017image, schmidt2020principles}.

Conventional dual-energy CT imaging methods are classified into three categories: projection-domain methods \cite{shi2019raw, stenner2007empirical}, image-domain methods \cite{dong2014combined, maass2009image, niu2014iterative, yu2012dual}, and one-step methods \cite{barber2016algorithm, mechlem2017joint, mory2018comparison, tilley2019model}. Except for the one-step method, both the projection-domain methods and the image-domain methods perform image reconstruction and material decomposition separately. The difference between the two methods is the order in which process is performed first. In contrast, the one-step methods literally perform reconstruction and material decomposition simultaneously.

In previous studies, similar to the conventional methods, several convolutional neural network (CNN) based methods have been suggested for material decomposition \cite{zhu2022feasibility, su2022direct}. As with the image-domain methods, some have applied CNNs for material decomposition in the image domain \cite{gong2020deep, zhang2019image}. Other CNN methods aim to achieve both image reconstruction and material decomposition in an end-to-end fashion like the one-step methods \cite{zhang2020dual}. By taking the advantages of including the dual-domain (the projection domain and the image domain) in the networks, the end-to-end CNNs have shown that their results are quantitatively superior to those of the conventional CNNs.

However, to our knowledge, none of the studies has discussed how to relax the data requirement of the DL approaches in DECT . The end-to-end network structures aggravate the massive data requirement because, in DECT, there are numerous combinations of inputs and outputs for the end-to-end network. Not only does the pair of X-ray tube energies for low-energy and high-energy change, but also the combination of basis materials varies, for example, iodine/tissue \cite{yamada2014radiotherapy}, bone/tissue \cite{schulz2012automatic}, liver tissue/fat \cite{walter2006accuracy}, xenon or krypton/lung tissue \cite{kong2014xenon}, etc.  Moreover, in clinical DECT, it is generally difficult to obtain the ground truth material-specific images for DL network training. Unlike the physical phantoms whose material information are exactly known, the clinical CT images would not provide accurate material maps of the human bodies. From dual-energy CT images or projections, one can only estimate the material-specific maps with its accuracy highly subject to the imaging conditions and the calibration methods. By the way, there is no widely agreed method or protocol for acquiring the ground truth images for DECT in clinical practice \cite{patino2016material}.

To overcome the challenge of data requirements across various energy pairs, we propose a novel training method that allows energy flexibility to a DL network in the energy range between two trained energies. Also, we limit the role of the proposed network only to image reconstruction. Thus, in our approach, once reconstruction is done by the proposed network, material decomposition would be conducted in the image domain by empirical methods, such as Ref. \cite{maass2009image, petrongolo2015noise}, which are much less burdensome than DL-based material decomposition methods in terms of data requirements.

In this paper, we propose a training method that consists of two techniques: standard-deviation normalization and energy mix-up. Through the training method, the network learns to reconstruct the image from the synthetic  data obtained at an energy between the two given energies. Furthermore, we propose a new CNN architecture inspired by the regularized version of the Primal-Dual Hybrid Gradient (PDHG) algorithm, which has an iterative dual-domain structure \cite{chambolle2018stochastic, chambolle2011first, valkonen2014primal}. One of the key features of the proposed network includes a subnetwork that is devised to take a regularizing role within the global network framework.

In this study, sparse-view DECT imaging scenario was adopted to demonstrate the feasibility of the proposed training method. Two simulated datasets based on 4D digital extended cardio torso (XCAT) phantom \cite{segars2008realistic, segars20104d} and 2-D high-resolution chest CT images \cite{sorensen2010quantitative} were used. The performance of the network is compared with other conventional algorithms and CNNs.

\section{Methods}

\subsection{\label{dblcol}Problem formation in DECT}

We formulate the linear inverse problem for reconstructing a ground truth image $\bold{f}^*\in \textrm{X}$ by assuming a monochromatic single-energy CT system as following:
\begin{equation}
    \bold{g}=\bold{Af^{*}}+\bold{\epsilon}
\end{equation}
where $g\in \textrm{Y}$ stands for the measured sinogram going through log transform, $A:\textrm{X}\rightarrow \textrm{Y}$ is the linear system matrix, and $\epsilon \in \textrm{Y}$ is noise for measurements. X and Y are the image space and the projection data space, respectively.
To solve ill-posed or severe noise-containing problems of (1), we conventionally deal with a model-driven optimization problem, and it is often expressed as following:
\begin{equation}
     \min_\bold{f} [L(\bold{Af},\bold{g})+R(\bold{f})] \quad\textrm{s.t.} \; \bold{f}\geq 0
 \end{equation}
 where $L$ is a loss function or a likelihood, and R is a prior-based regularization term. In DECT, there exists each optimization problem for each energy spectra. Here, we express the optimization problem for DECT image reconstruction by using an additional energy index, $e$, as follows:
 \begin{equation}
     \min_\bold{f} [L(\bold{Af^{e}},\bold{g^{e}})+R(\bold{f^{e}})] \quad\textrm{s.t.} \; \bold{f}\geq 0, e \in\ [E_{1},E_{2}]
 \end{equation}
 
In CT imaging, the above optimization problem has been often resolved by gradient-based optimization methods, such as gradient descent \cite{sidky2008image}, conjugate gradient methods \cite{kawata1985constrained}, Quasi-Newton methods \cite{fessler1997grouped, elbakri2003segmentation, kim2014sparse}, etc.

  As will be detailed in the following, we use a primal-dual algorithm with its functions replaced by a deep-learning network in this work. 

\begin{figure}
    \centering
    \includegraphics[scale=.6]{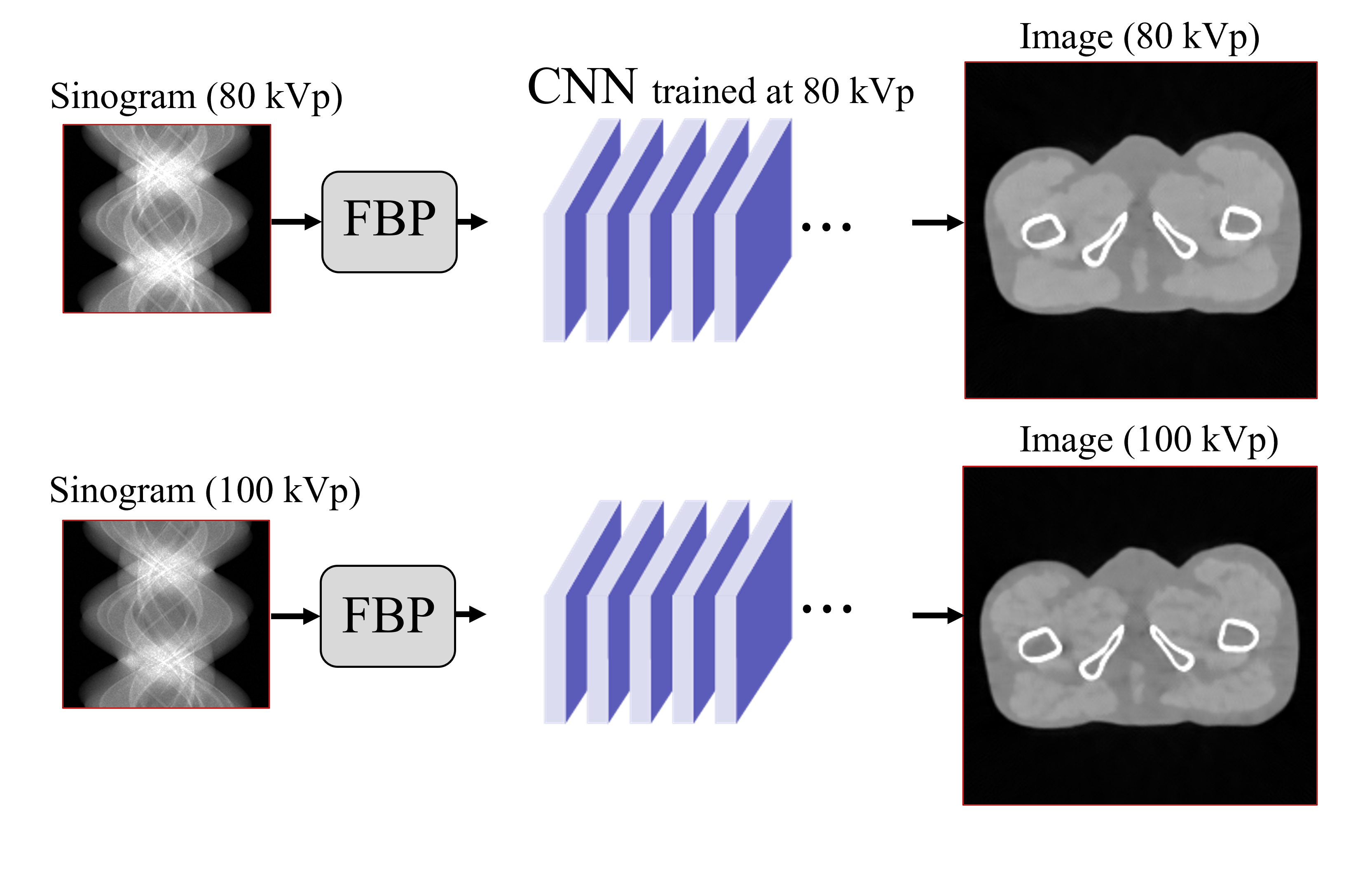}
    \caption{An example CNN that is not trained to be energy-flexible and its results for two different X-ray tube energies. In this example, U-Net [19] trained by 80kVp data is used. Compared to the 80kVp image, soft tissues in the 100kVp image are severally contaminated by artifacts.}
    \label{fig1}
\end{figure}

\subsection{Energy Flexibility Training (EFT) Method }

 Most of the earlier proposed DL networks have been conventionally trained for data or images acquired at the specified X-ray energy; thus, their performances at other X-ray energies were not guaranteed as an example shown in Figure \ref{fig1}. The CNN in this case was used for denoising, but it failed to denoise a CT image acquired at another tube voltage. 

In this study, we propose a training method, which extends the given data space, to make a neural network energy-flexible within two trained X-ray energies. Here, energy flexibility stands for the property of a network or an algorithm that is robust against various X-ray tube energies. For example, one can say that the filtered back-projection (FBP) is one of the energy-flexible operators because it contains no energy-dependent parameter, and the performance is consistent for any X-ray energy setting \cite{feldkamp1984practical}.

The proposed method, named Energy Flexibility Training (EFT) Method, consists of two techniques. The first technique is a standard deviation normalization (SN) in the projection data and a standard deviation denormalization (SD) in the reconstructed image. Those are expressed as the following equations:
\begin{equation}
\begin{array}{l}
SN(\bold{g})=\sigma_{\bold{A}^{T}g}^{-1}\bold{g} \\
SD(\bold{f})=\sigma_{\bold{A}^{T}g}\bold{f}
\end{array}
\end{equation}
where $\sigma_{\bold{A}^{T}\bold{g}}$ and $\sigma_{\bold{A}^{T}\bold{g}}^{-1}$ are the standard deviation over the entire reconstructed image, $\bold{A}^{T}\bold{g}$, and its reciprocal, respectively. $\bold{A}^{T}$ is implemented by FBP with the Ram-Lak filter. The standard deviation of the reconstructed image has a linear relationship with that of the measured sinogram because of the linearity of FBP, i.e., $\sigma_{\bold{A}^{T}(c\bold{g})}=c\sigma_{\bold{A}^T\bold{g}}$ for $\forall c\in\mathbb{R}^{+}$. By being included in both ends of a network, the SN and SD play a role in relaxing the scale dependency of a network on the measured sinogram that enters the network as an input and returning the scale back to the network's output, respectively. We will explain how the SN and SD are connected to and used in the proposed network in Section 2.3. 

The second training technique is an energy mix-up. Mix-up is one of the data augmentation methods proposed in the previous publications on DL network training \cite{he2019bag, zhang2017mixup}. By interpolating training data, one can extend the given data space and improve the generalizability of neural networks. In this study, we implemented the mix-up between two different X-ray energy data thus named energy mix-up. The energy mix-up generates virtual data and images that are linearly mixed between those at two X-ray energies:
 \begin{equation}
 \begin{array}{l}
 \bold{\tilde{g}}=\lambda \bold{g}^{e_{1}}+(1-\lambda) \bold{g}^{e_{2}} \\
 \bold{\tilde{f}}^{*}=\lambda \bold{f}^{e_{1}*}+(1-\lambda) \bold{f}^{e_{2}*}
 \end{array}
 \end{equation}
 where $\lambda\in [0,1]$. $\bold{f}^{e_{1}*}$ and $\bold{f}^{e_{2}*}$ are the ground truth reconstructed images at the energy $e_{1}$ and $e_{2}$, respectively. Different with the SN and SD, the energy mix-up is only conducted in the pre-processing stage of training data. For each training data at each training epoch, the parameter $\lambda$ in equation (5) is randomly selected in the range of [0,1] and applied.

\begin{figure}
    \centering
    \includegraphics[scale=.5]{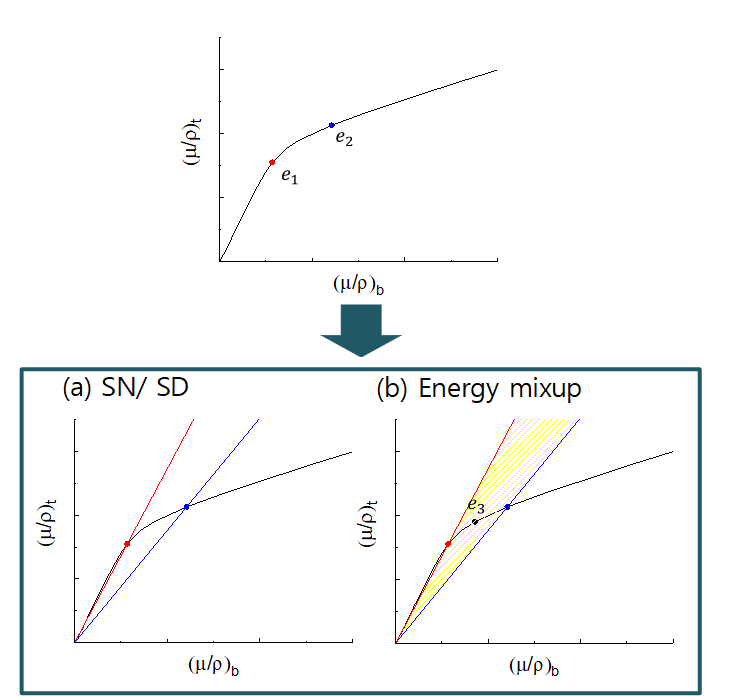}
    \caption{Graph description of the role of the proposed EFT. The colored points, lines, and area from the top to the bottom right image visually show the expansion effect of each proposed training technique on the data space.}
    \label{fig2}
\end{figure}
 
 For a conceptual illustration of the effects of the proposed EFT, let us suppose an ideal situation where the X-ray energy is monochromatic and the imaged object is of two components, i.e., soft-tissue and bone. In a 2D data space where each dimension represents one of the mass attenuation coefficients of the components, which is an upstream space of all measurements, we can represent a CT scan as a point. CT scans with two different energies would be two points such as $e_1$ and $e_2$ on the first graph in Figure \ref{fig2}. As the energy of X-ray varies, the corresponding point would also be displaced as shown in Figure \ref{fig2} tracing an implicitly underlying nonlinear curve. If a network is trained only by the data that belong to the two points in DECT, the network performance would be subject to the X-ray energy corresponding to the test data. By adding the SN and SD to both ends of the network, one can extend the data space from the two points to two lines connecting those points from the origin, as shown in Figure \ref{fig2}(a). It is because the standard deviation normalization makes all mass attenuation points from a line the same. Furthermore, the energy mix-up again expands the data space from the two lines to a planar area closed by those lines. The expanded data space would densely fill the area where other X-ray energy data points between the two trained X-ray energies, such as $e_{3}$ in Figure \ref{fig2}(b), are included. It should be noted that the effect of extending the data range through the EFT as above does not mean an appropriate choice of $\lambda$ would allow us to synthesize physically measurable data or corresponding images at an arbitrary energy between the two. Instead, the EFT extends given data space so that it densely includes synthetic data points around unknown data points between the two trained energies.

\begin{figure*}
    \centering
    \includegraphics[scale=0.55]{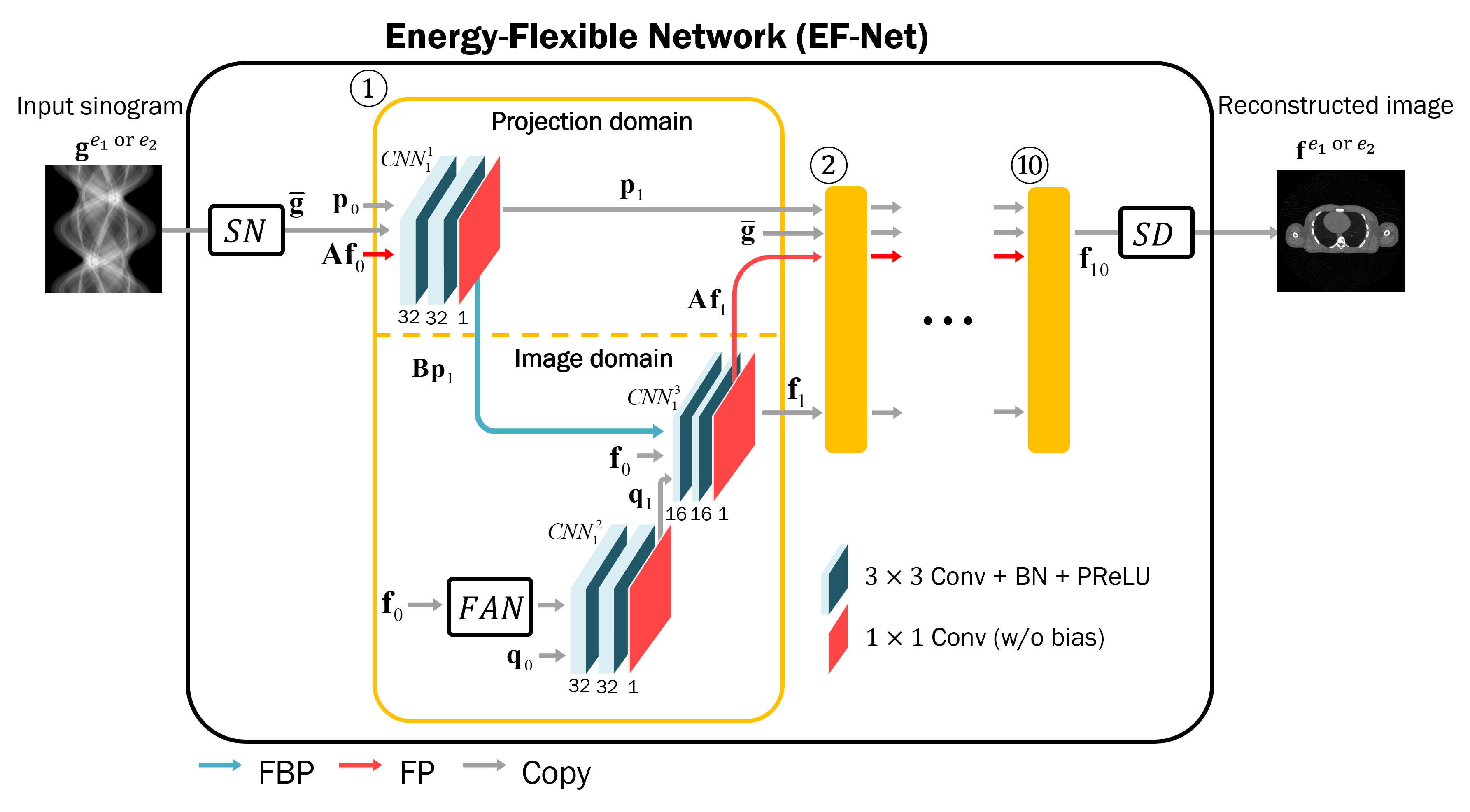}
    \caption{Energy-Flexible Network (EF-Net). The network contains ten yellow boxed subnetworks. The colored three-dimensional boxes represent the components of the CNNs in Algorithm \ref{alg2}. The small numbers above or under the three-dimensional boxes represent the number of output channels. The colored arrows between the boxes stand for FBP, FP, or copy. If multiple arrows point to a box together, it means that they are concatenated in the channel direction and then fed to the box. For the FBP, the Ram-Lak filter is used.}
    \label{fig3}
\end{figure*}

 \subsection{Proposed Network}
 The primal-dual algorithms have several benefits over other optimization methods. One of the key benefits is that the primal-dual algorithms allow us to generalize the optimization problems rather easily in various forms. In this work, we used the PDHG algorithm, which is one of the widely used primal-dual algorithms. The following optimization problem is substituted for (2) in the PDHG algorithm:
 \begin{equation}
      \min_{\bold{f}^{e}} [F_{1}(K_{1}(\bold{f}^{e}))+F_{2}(K_{2}(\bold{f}^{e}))+G(\bold{f^{e}})]
 \end{equation}
 where we define $F_{1}(\cdot)\equiv L(\cdot, \bold{g}^e)$, and $K_{1}\equiv \bold{A}$. $G(\cdot)$ is a positive indicator function that serves as the non-negative constraint. Here, $R(\bold{f}^e)$ in (3) is expressed in the combination of a sparsifying transform operator, $K_2$, and its loss function, $F_2$, i.e. $F_2(K_2)\equiv R$. For example, total variation (TV) regularization can be interpreted as the combination of the gradient function, $K_2=\bigtriangledown$, and its $l_1$ loss function, $F_2(\cdot)=\beta $. The pseudo-code to solve the regularized version of the PDGH algorithm in (6) is given in Algorithm \ref{alg1}. In Algorithm \ref{alg1}, $F_1^*$ and $F_2^*$ refer to the Fenchel conjugate of $F_1$ and $F_2$, respectively, and $K=(K_1,K_2)$. $\bold{B}$ is a pseudo-inverse of $\bold{A}$, and it refers to the FBP in the CT image reconstruction. $\bold{p}$ and $\bold{q}$ are vectorized image arrays having the same size with $\bold{g}$ and $\bold{f}^e$, respectively. 

\begin{algorithm}
\caption{PDHG}\label{alg1}
\begin{algorithmic}[1]
\State $\sigma,\tau >0 \quad s.t.\; \sigma\tau <1, \theta\in[0,1]$
\State $\bold{f}_{n}^e, \bold{\overline{f}}_{n}^e, \bold{q}_n \in X, \bold{g}^e,\bold{p}_n \in Y$
\For{$n=0,N$}
    \State $\bold{p}_{n+1} \leftarrow \textrm{prox}_{\sigma}[F_{1}^*](\bold{p}_n+\sigma K_1(\bold{\overline{f}}_n^e))$
    \State $\bold{q}_{n+1} \leftarrow \textrm{prox}_{\sigma}[F_{2}^*](\bold{q}_n+\sigma K_2(\bold{\overline{f}}_n^e))$
    \State $\bold{f}_{n+1}^e \leftarrow \textrm{prox}_{\tau}[G](\bold{f}_n^e-\tau K^T(\bold{p}_{n+1}+\bold{q}_{n+1})$
    \State $\bold{\overline{f}}_{n+1}^e \leftarrow \bold{f}_{n+1}^e + \theta (\bold{f}_{n+1}^e - \bold{f}_{n}^e)$
\EndFor
\State $\bold{return} \;\bold{\overline{f}}_{N}^{e*}$
\end{algorithmic}
\end{algorithm}

The proposed network is largely based on Algorithm \ref{alg1}. It is briefly summarized in Algorithm \ref{alg2}. We replace each update step of Algorithm \ref{alg1} by the subnetwork in which three convolutional layers with batch normalization layers and activation operators are included. The detailed architecture of the proposed network is depicted in Figure \ref{fig3}. All the elements of the initial image arrays, $\bold{\overline{f}}_{0}^{e*}$, $\bold{q}_0$, and $\bold{p}_0$, are first initialized by zero, and then the measured sinogram $\bold{g}^e$ is normalized by SN. The normalized sinogram $g^e$ is first fed to $CNN_n^1$ with $\bold{p_0}$ and $\bold{A}\bold{\overline{f}}_{0}^{e*}$. At the same time, $f_0^e$ is fed to $CNN_n^2$ with $\bold{q}_0$. Next, the filtered back-projection output of $CNN_n^1$ and the output of $CNN_n^2$ are concatenated and fed to $CNN_n^3$ with $\bold{\overline{f}}_n^e$. All the outputs of the subnetworks become the inputs in the next iteration. These steps are repeated ten times. Lastly, the last iteration output $\bold{\overline{f}}_{10}^e$ is denormalized by SD and returns the network output  $\bold{f}_{N}^{e*}$.

\begin{figure}
    \centering
    \includegraphics[scale=.6]{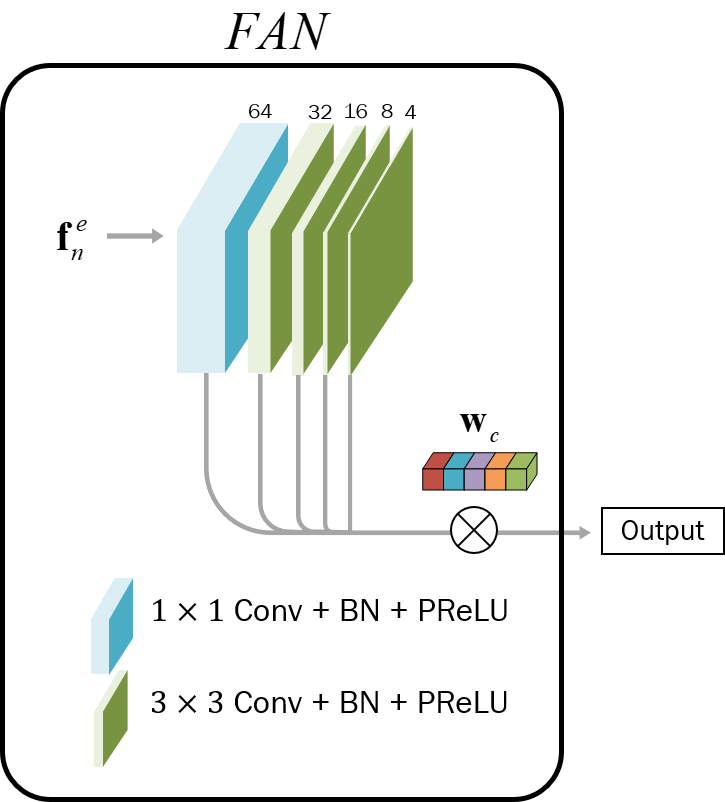}
    \caption{Architecture of the subnetwork, Frequency Attention Network (FAN). The input $\bold{f}_n^e$ is fed to colored CNN layers, and the output of the network is the sum of the CNN layers channel-directionally weighted by the $\bold{w}_c$.}
    \label{fig4}
\end{figure}

Developing CNNs based on the PDHG algorithm have been already tried \cite{chen2021deep, li2020proximal, meinhardt2017learning}. However, one distinctive feature of the proposed network compared to the previous PDHG-based CNNs is the Frequency Attention Network (FAN) and the second subnetwork, $CNN_n^2$. The FAN and the second subnetwork replace the roles of the sparsifying transform operator, $K_2$, and its loss function, $F_2$, in (6), respectively. They are distinctly connected to the global network and only deal with $\bold{\overline{f}}_n^e$ as the input. The architecture of the FAN is shown in Figure \ref{fig4}. Inspired by regularizations based on frequency weighting in the iterative algorithms \cite{lee2017feasibility, sidky2011constrained}
, the architecture of the FAN is designed to gradually focus from high to low-frequency information of the image. The front CNN layers of the network relatively focus on high-frequency information. The rear CNN layers focus on low-frequency information because elements in a wider area are incorporated. All the outputs of the layers are then concatenated and summed up after multiplying the trainable channel directional weights, $\bold{w}_c$. Similar to the sparsifying transform operator in the iterative algorithm that is conventionally treated as a deterministic operator based on prior knowledge of the desired image, the FAN also recurs by using the same parameters throughout the iterations in the network.

\begin{algorithm}
\caption{The proposed network}\label{alg2}
\begin{algorithmic}[1]
\State $\bold{\overline{f}}_{n}^e, \bold{q}_n \in X, \bold{g}^e,\bold{p}_n \in Y$
\State $\bold{\overline{g}}_{n}^e = SN(\bold{f}_{n}^e)$
\For{$n=0,N$}
    \State $\bold{p}_{n+1} \leftarrow CNN_n^1(\bold{p}_n,\bold{A}\bold{\overline{f}}_{n}^e,\bold{\overline{g}}_n^e)$
    \State $\bold{q}_{n+1} \leftarrow CNN_n^2(\bold{q}_n,FAN(\bold{\overline{f}_n^e}))$
    \State $\bold{\overline{f}}_{n+1}^e \leftarrow CNN_n^3(\bold{\overline{f}}_{n}^e, \bold{B}\bold{p}_{n+1},\bold{q}_{n+1})$
\EndFor
\State $\bold{f}_{N}^{e*} = SD(\bold{\overline{f}}_{N}^e)$
\State $\bold{return} \;\bold{f}_{N}^{e*}$
\end{algorithmic}
\end{algorithm}

\subsection{Data Preparation}

\begin{figure}
    \centering
    \includegraphics[scale=.6]{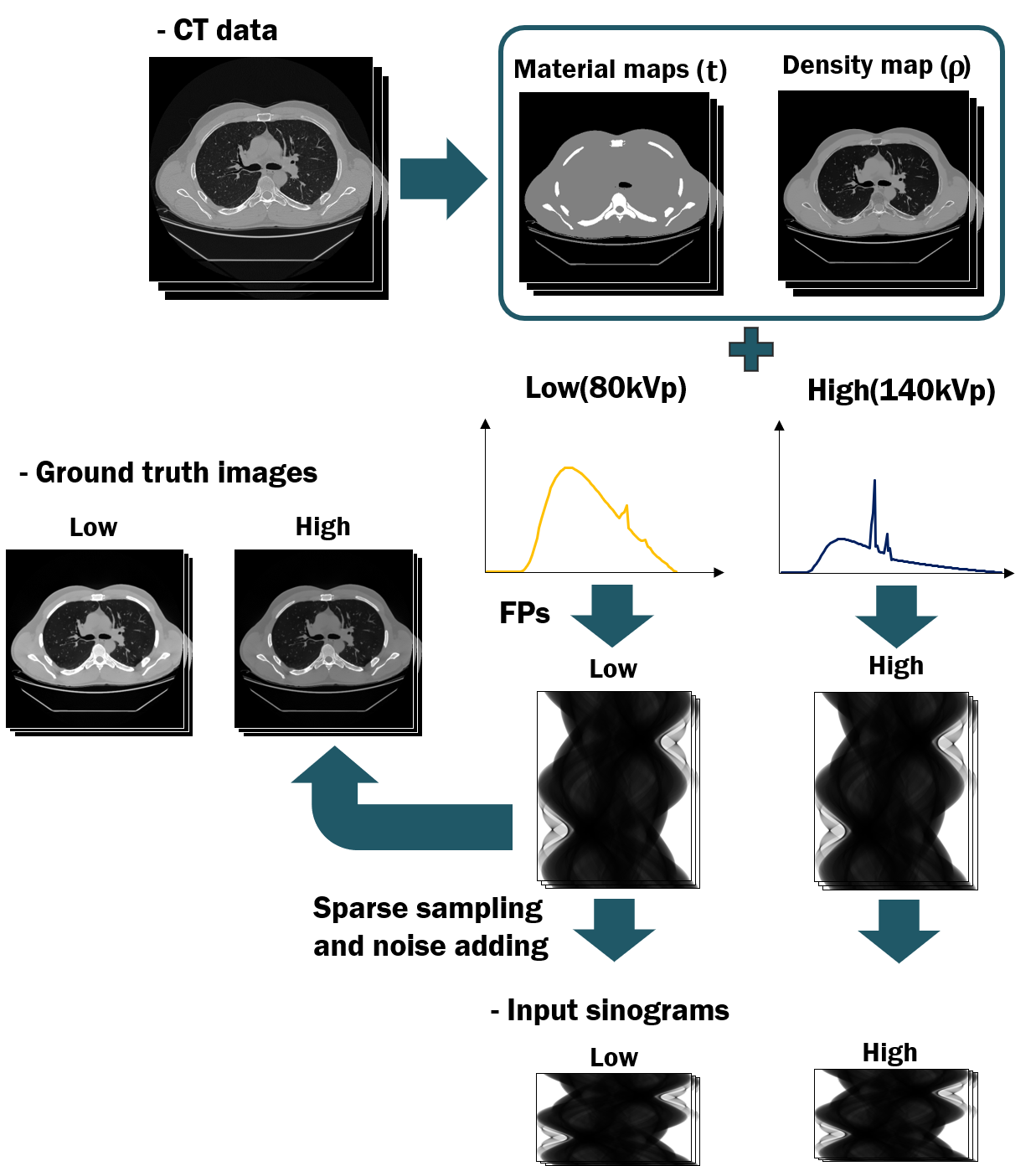}
    \caption{Workflow of the simulation data generation.}
    \label{fig5}
\end{figure}

We tested the proposed algorithm with two simulated datasets: the XCAT phantom dataset \cite{segars20104d} and the clinical chest CT image dataset \cite{sorensen2010quantitative}. Both datasets were simulated for fan-beam CT scans and generated by following the workflow illustrated in Figure \ref{fig5}.

First, we segmented the CT images into four (the XCAT phantom data) or two (the clinical chest CT data) materials to generate material maps and density maps. The material maps, $\bold{t}$, are the segmented CT images having material indices. To generate the density map, $\boldsymbol{\rho}$, we first set the reference Hounsfield unit (HU) and density for each material (e.g., 25 HU and 1.04  $\sfrac{g}{m^3}$ for soft tissue). Then, for each pixel, we multiplied the CT pixel value by the corresponding reference density, and then divide it by the reference HU. The calculated value for each pixel represents the density of the corresponding material on the density map. 

 Next, polychromatic sinograms were generated by performing the forward projection at the given x-ray spectra. For both datasets, we simulated the fan-beam CT system in which the source-to-detector distance and the source-to-isocenter distance are 1300mm and 900mm, respectively. The attenuation coefficient at the $k^{th}$ image pixel for the $j^{th}$ energy bin is calculated as
  \begin{equation}
f_{jk}^{'} = \boldsymbol{\rho}(\frac{\mu}{\rho_{m}})_{j,\bold{t}_k}
 \end{equation}
 where $(\sfrac{\mu}{\rho_{m}})_{j,\bold{t}_k}$ is the mass attenuation coefficient of the material at the $k^{th}$  image voxel and at the $j^{th}$ energy bin. We obtained the mass attenuation coefficient in the National Institute of Standard Technology database (NIST) \cite{seltzer1995tables}. The X-ray spectra used in this simulation were obtained from the open-source X-ray spectrum simulator \cite{despres2013spectrum}.

Finally, for both datasets, we used uniformly sampled 180 and 720 views over 360 degrees for the sparsely sampled input data and the ground truth images, respectively. The sinograms for all energy spectra were obtained in a view-matched manner in terms of the source and detector position. Poisson noise was added to the input sinograms, but the ground truth images were reconstructed from the noise-free 720-view projection data. The FBP with the Hamming filter was used for reconstructing the ground truth images.

\textbf{1) an XCAT phantom dataset:} The XCAT phantom was set to contain a total of four materials: bone and three types of tissue (adipose tissue, soft tissue, and lung tissue). The XCAT phantom data have relatively simple and plain structures, and the density for each material is predefined. Among the total 600 XCAT phantom slice images from the pelvis to the head, we randomly selected and used 580 slices for training, 10 for validation, and 10 for testing. The sinograms were simulated to have 256 detector pixels with a 1.6 mm pixel pitch. For the sinogram, we equally added Poisson noise corresponding to $10^3$  incident photons per pixel for all energy data, which is approximately equivalent to $3.16\%$ Gaussian noise. The reconstructed image size is  256×256.

\textbf{2) a chest CT image dataset:} The clinical chest CT images have relatively complicated structures, and the density of each material substantially varies. We segmented the CT image into soft tissue and bone based on the CT numbers. The dataset includes in total 112 high-resolution CT slice images, coming from 39 patients. Among them, randomly selected 96, 8, and 8 slices were used for training, validation, and testing, respectively. For this simulation, we obtained the sinogram with 512 pixels having a 0.8 mm pixel pitch to better represent the spatial resolution of the original images. We added Poisson noise corresponding to $2*10^4$  incident photons per pixel for all energy data, which is approximately equivalent to $0.7\%$ Gaussian noise. The reconstructed image array size is the same as the XCAT simulation.

\subsection{Network Training Details}

All the parameters over ten iterations in the proposed network are simultaneously updated by minimizing the mean-square-error between the output of the proposed network and the ground truth images as below: 

\begin{equation}
     \min_\bold{\Theta} L(\bold{D};\bold{\Theta}) = \frac{1}{N_{\bold{D}}}\sum_{d=1}^{N_{\bold{D}}} \| E(\bold{\tilde{p}}_d;\bold{\Theta})-\bold{\tilde{f}}_{d}^{*} \|_{2}^{2}
\end{equation}
where $\bold{D}$ is the training dataset, and $N_{\bold{D}}$ is the number of the training dataset. $\bold{\Theta}$ refers to a set of parameters of the proposed network and $E$ is the proposed network operator depicted in Fig. 3. The optimal iteration number in iterative networks is subject to many factors; thus, we used the fixed number of iterations, rather than conducting an exhaustive search for the optimal iteration number, which is computationally expensive. The iteration number, 10, provides proper benefits of the iteration and still shows faster computational time for reconstruction. 

The implementation of the proposed network was done with TensorFlow \cite{abadi2016tensorflow}. The forward and backward projections in the EF-Net were implemented by using the tomography-related operators and the TensorFlow dedicated layer transformation function in Operator Discretization Library (ODL) \cite{adler2017operator}. The Adam optimization algorithm was used for training \cite{kingma2014adam}. For the initial learning rates, we used $10^{-2}$ for the X-CAT phantom data and $10^{-3}$ for the chest CT data, which are exponentially decaying with the rate $\beta_1=0.99$ and $\beta_2=0.999$. All parameters in the proposed method were initialized by the Xavier initializer in TensorFlow, and the initial reconstructed image, $f_0$, was initialized by zero \cite{glorot2010understanding}.

The computing resources used for training include an Intel® Core i9-10900X CPU and a NVIDIA GeForce RTX 2080Ti GPU. The EF-Net was trained for 4k and 11k epochs for the XCAT phantom data and the chest CT data, respectively.

\subsection{Comparison Algorithms}
We compared our proposed network with other algorithms that are well known or related to the sparsely sampled CT image reconstruction.

\textbf{1) FBP:} the FBP using the Hamming filter was used to show the basic image quality for given sparse, noisy measurements.

\textbf{2) TV-GMI:} This is the iterative reconstruction algorithm that effectively improves the image quality by exploiting the gradient-based joint sparsity between two energy images as the regularization \cite{lee2017feasibility, cho2020novel, zhang2021directional}.

\textbf{3) U-Net:} The U-Net is a widely used DL network in the medical imaging fields \cite{jin2017deep, ronneberger2015u}. The U-Net efficiently treats the multi-frequency information of input data by using the max-pooling layers and the up-convolution layers in the image domain. In this study, a reconstructed image by the FBP was fed as an input to the U-Net having a skip connection.

\textbf{4) Learned primal-dual (LPD):} The learned primal-dual algorithm is the DL network that originates from the PDHG   algorithm, as with the proposed algorithm \cite{adler2018learned}. We compared our proposed algorithm with the learned primal-dual algorithm to show how the proposed schemes work in contrast. For a fair comparison, the same iteration number, 10, as the proposed network was used.

The two DL networks were basically trained in the same way as the proposed network. The same loss function and optimization algorithm for training were used, and the initial step size for each network for each dataset was empirically chosen.

\begin{figure}
    \centering
    \includegraphics[scale=1.0]{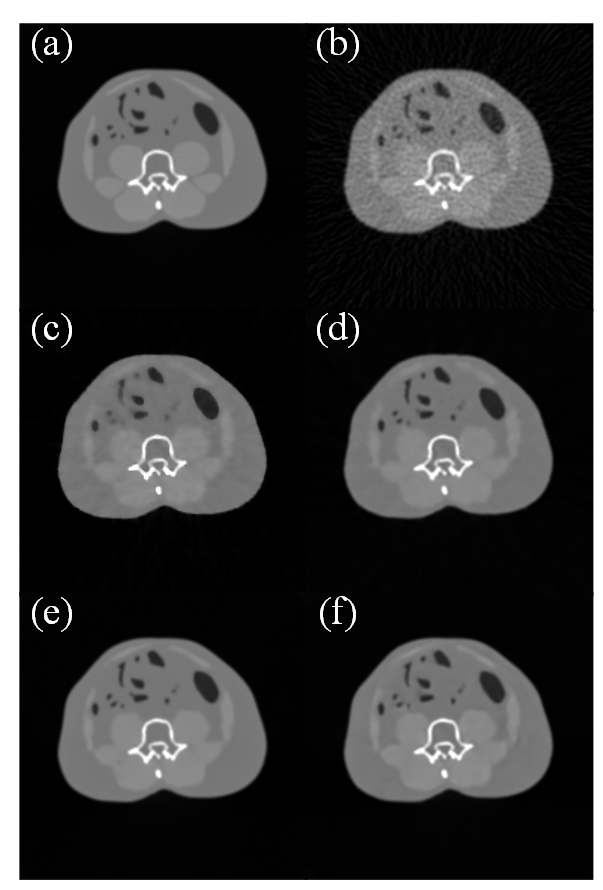}
    \caption{Reconstructed images of the 140kVp X-CAT phantom test data are shown. (a) Ground truth, (b) FBP, (c) TV-GMI, (d) U-Net, (e) LPD, and (f) EF-Net. Display window: [0, 0.04]  $\textrm{mm}^{-1}$}
    \label{fig6}
\end{figure}

\section{Results}
For two simulation datasets, we evaluated the performance of the algorithms on both trained energy data and untrained energy data. Specifically, we used data acquired at X-ray tube energy 80kVp and 140kVp for the training and used data acquired at the tube voltages of 100kVp and 140kVp for testing. The 140kVp test data were used to check the conventional reconstruction performance of the networks when training data and test data are matched with respect to the X-ray tube voltage, while the 100kVp test data were used to test the energy flexibility of the networks.

\begin{figure}[!t]
    \centering
    \includegraphics[scale=1.0]{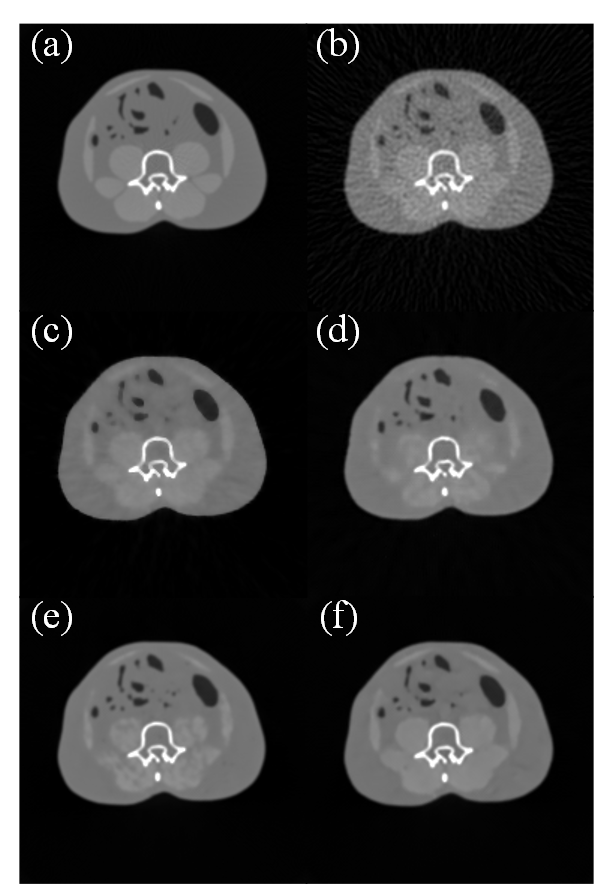}
    \caption{Reconstructed images of the 100kVp X-CAT phantom test data are shown. (a) Ground truth, (b) FBP, (c) TV-GMI, (d) U-Net, (e) LPD, and (f) EF-Net. Display window: [0, 0.045]  $\textrm{mm}^{-1}$}
    \label{fig7}
\end{figure}

\subsection{XCAT phantom study}
We first compared the performance of all the algorithms for 140kVp test data. U-Net and LPD were trained by 140kVp training data only. In contrast, EF-Net was trained using both 80kVp and 140kVp data with the EFT. For TV-GMI, two hyperparameters for regularization terms were empirically determined. Figure \ref{fig6} shows the reconstructed images by all the algorithms along with the ground-truth image. The TV-GMI effectively reduces noise shown in the image of FBP, but the image quality of the TV-GMI is visually inferior to those of the three DL-based methods. The LPD and the EF-Net show the largest improvement among them. The result of the U-Net looks a little blurrier than those of the LPD and the EF-Net.

\begin{table*}
\caption{\label{tab1}Quantitative results of the reconstruction methods for the 140kVp XCAT phantom test data including the number of parameters and runtime.}
\centering
\begin{indented}
\item[]\begin{tabular}{@{}*{6}{l}} 
 \br
 Methods & RMSE (HU) & SSIM & Parameters & Runtime (s) \cr
 \mr
 FBP & 57.93 & 0.8844 & $\boldsymbol{1}$ & $\boldsymbol{0.35}$ \cr
 TV-GMI & 35.19 & 0.9780 & 2 & 20.4 \cr
 U-Net & 20.28 & 0.9840 & $9.67\times 10^6$ & 1.12 \cr
 LPD & 13.35 & 0.9930 & $2.53\times 10^5$ & 0.74 \cr
 EF-Net & $\boldsymbol{13.25}$ & $ \boldsymbol{0.9938}$ & $2.58\times 10^5$ & 1.30 \cr 
 (w/o EFT) & & & & & \cr
 EF-Net & 14.92 & 0.9925 & $2.58\times 10^5$ & 1.33 \cr
 \br
\end{tabular}
\end{indented}
\end{table*}

\begin{table}
\caption{\label{tab2}Quantitative results of the reconstruction methods for the 100kVp XCAT phantom test data.}
\centering
%\begin{indented}%
\begin{tabular}{@{}*{4}{l}} 
 \br
 Methods & RMSE (HU) & SSIM & \cr
 \mr
 FBP & 62.80 & 0.8749 \cr
 TV-GMI & 35.20 & 0.9782 \cr
 U-Net & 25.01 & 0.9799 \cr
 LPD & 20.01 & 0.9828 \cr
 EF-Net & 19.13 & 0.9841 \cr 
 (w/o EFT) & & & \cr
 EF-Net & $\boldsymbol{15.67}$ & $\boldsymbol{0.9916}$ \cr
 \br
\end{tabular}
%\end{indented}%
\end{table}

\begin{figure*}[!t]
    \centering
    \includegraphics[scale=.4]{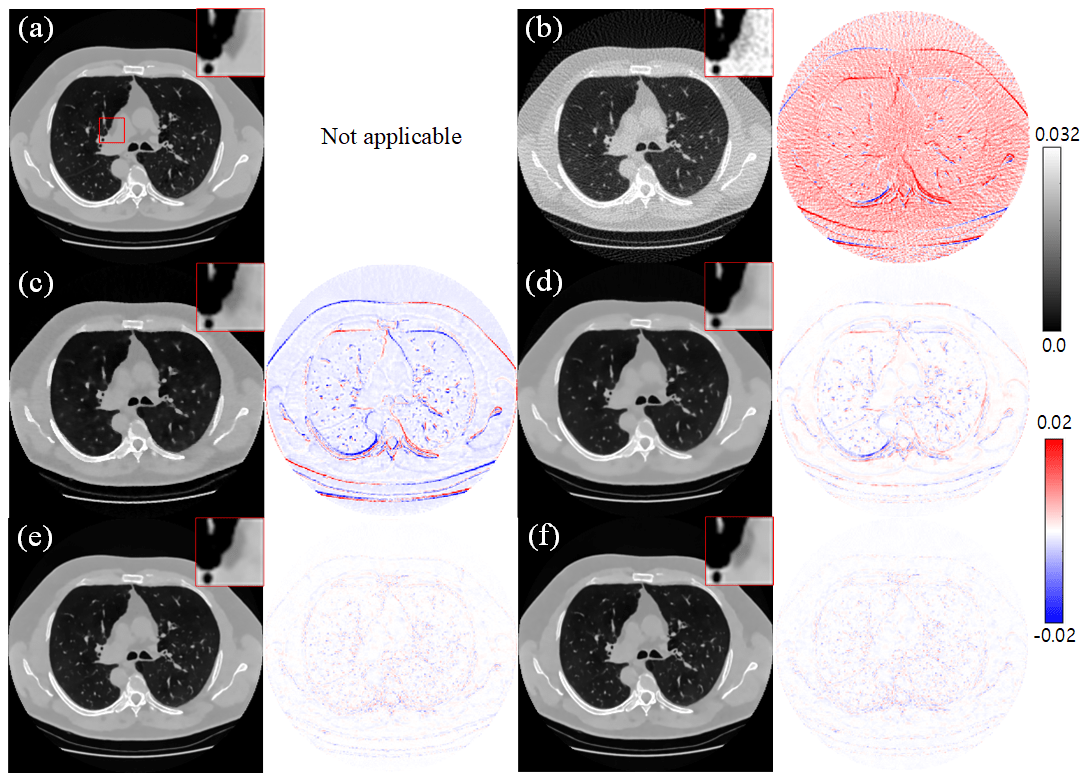}
    \caption{Reconstructed images of the 140kVp clinical chest CT test data are shown, and their error maps are represented below the corresponding images. (a) Ground truth, (b) FBP, (c) TV-GMI, (d) U-Net, (e) LPD, and (f) EF-Net. Display windows for the CT images and the error maps are attached.}
    \label{fig8}
\end{figure*}

We also conducted a quantitative analysis for all the test data and took averages of the results. In this study, two metrics are chosen for quantitative analysis: Root-Mean-Square Error (RMSE) and Structural Similarity Index Measure (SSIM). The quantitative results and runtimes for the 140kVp dataset are listed in Table \ref{tab1}. In the quantitative analysis, we additionally included the quantitative results of EF-Net trained without using the EFT to test the original performance of the proposed network performance, and it is named as ‘EF-Net (w/o EFT)’ in the table. The result of the EF-Net without the EFT was the best in terms of the two metrics, and the LPD was the second-best. The EF-Net was the next with similar SSIM to the EF-Net without the EFT and the LPD. For runtime, the TV-GMI is the slowest and LPD is the fastest among the algorithms.

The next comparison was conducted with the 100kVp test data. In this case, U-Net, LPD, and EF-Net without the EFT were trained by 80kVp data, while we used the same EF-Net trained earlier with both 80kVp and 140kVp data. Figure \ref{fig7} shows the reconstructed images. The images of the FBP and the TV-GMI have visually similar image quality to their 140kVp results. Also, the EF-Net shows a similar degree of image improvement. On the other hand, the U-Net and the LPD produce severe artifacts, especially in the soft tissue regions. 
Quantitative results for the 100kVp images are summarized in Table \ref{tab2}. The quantitative results of the U-Net, the LPD, the EF-Net without using the EFT are substantially worse than those of the 140kVp results. Among all the deep learning networks, the EF-Net shows the most noticeable improvement over the FBP and the TV-GMI.

\begin{figure*}[!t]
    \centering
    \includegraphics[scale=.4]{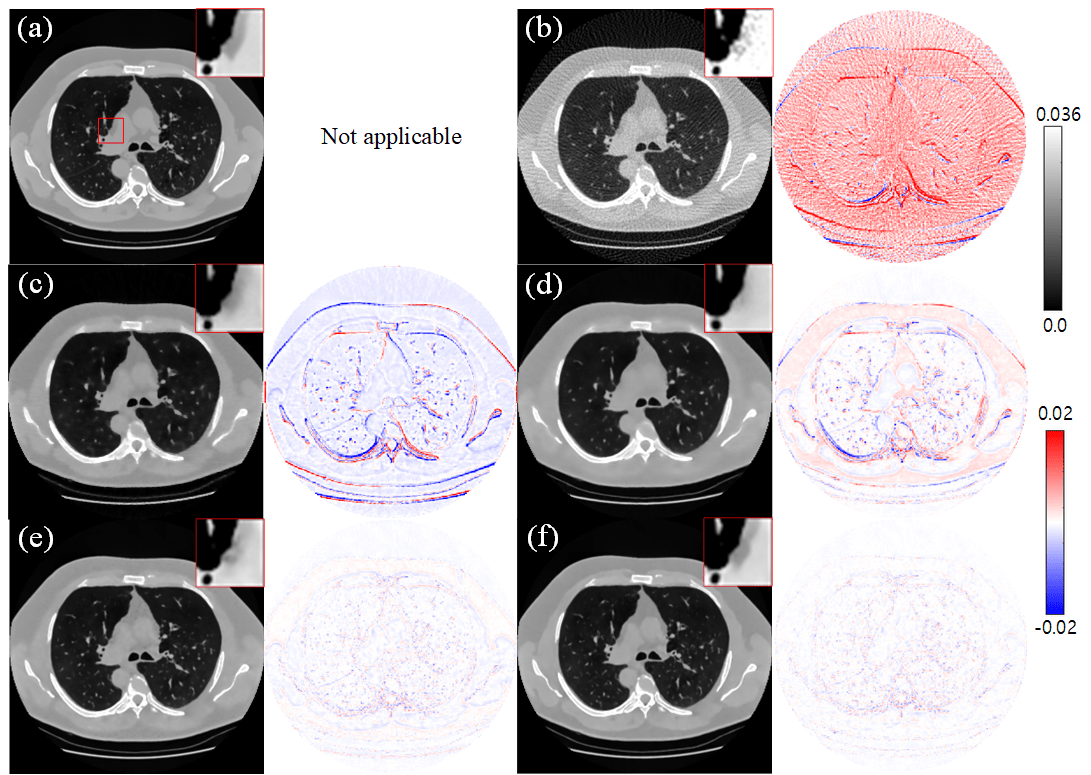}
    \caption{Reconstructed images of the 100kVp clinical chest CT test data are shown, and their error maps are represented below the corresponding images. (a) Ground truth, (b) FBP, (c) TV-GMI, (d) U-Net, (e) LPD, and (f) EF-Net. Display windows for the CT images and the error maps are attached.}
    \label{fig9}
\end{figure*}

\subsection{Chest CT study}
We tested the performance of all the methods with respect to the chest CT data. Figure \ref{fig8} shows the reconstructed images from one of the 140kVp test data. On the right side of each reconstructed image, we present an error map, which was made by subtracting the ground truth image from the reconstructed images. The FBP result seems to be noisy and overestimated due to the sparse-view artifacts while the TV-GMI result is over-smoothed and underestimated. Similar to the previous simulation, the U-Net shows much improvement, but the result is relatively over-smoothed, especially around the edges of structures. The error maps clearly show these differences among the reconstruction methods. The LPD and EF-Net results are visually very similar to each other and have relatively small errors with respect to the ground truth.

 Quantitative results of the 140kVp chest CT reconstructed images are summarized in Table \ref{tab3}. Different from the previous XCAT results, both the EF-Net without the EFT and the EF-Net show better improvement over the FBP than the LPD in terms of the two metrics. The runtimes for all the algorithms were slightly longer than those of the XCAT data overall. It is thought to be due to the larger size of the chest CT sinograms.

\begin{table*}
\caption{\label{tab3}Quantitative results of the reconstruction methods for the 140kVp chest CT test data including the number of parameters and runtime.}
\centering
%\begin{indented}
\begin{tabular}{@{}*{6}{l}} 
 \br
 Methods & RMSE (HU) & SSIM & Parameters & Runtime (s) \cr
 \mr
 FBP & 143.9 & 0.6562 & $\boldsymbol{1}$ & $\boldsymbol{0.42}$ \cr
 TV-GMI & 85.09 & 0.9272 & 2 & 25.0 \cr
 U-Net & 38.14 & 0.9745 & $9.67\times 10^6$ & 1.34 \cr
 LPD & 17.97 & 0.9910 & $2.53\times 10^5$ & 0.95 \cr
 EF-Net & $\boldsymbol{17.06}$ & $ \boldsymbol{0.9930}$ & $2.58\times 10^5$ & 1.72 \cr 
 (w/o EFT) & & & & & \cr
 EF-Net & 17.58 & 0.9925 & $2.58\times 10^5$ & 1.76 \cr
 \br
\end{tabular}
%\end{indented}
\end{table*}

\begin{table}
\caption{\label{tab4}Quantitative results of the reconstruction methods for the 100kVp chest CT test data.}
\centering
%\begin{indented}
\begin{tabular}{@{}*{4}{l}} 
 \br
 Methods & RMSE (HU) & SSIM \cr 
 \mr
 FBP & 150.0 & 0.6536 \cr
 TV-GMI & 85.78 & 0.9261 \cr
 U-Net & 48.26 & 0.9395 \cr
 LPD & 21.16 & 0.9865 \cr
 EF-Net & 21.12 & 0.9872 \cr 
 (w/o EFT) & & & \cr
 EF-Net & $\boldsymbol{19.01}$ & $\boldsymbol{0.9916}$ \cr
 \br
\end{tabular}
%\end{indented}
\end{table}

 The reconstructed images and the quantitative results for the 100kVp chest CT data set are shown in Figure \ref{fig9} and Table \ref{tab4}. Again, U-Net and LPD were trained by 80kVp training data, but we used the same EF-Net that was trained by both 80kVp and 140kVp training data. Like the previous XCAT results, soft tissue regions for the U-Net and the LPD are severely degraded. Compared to the performance degradation of the two, the U-Net performance drop is worse. The quantitative results of the EF-Net without the EFT is still slightly better than the LPD but similarly worse than its 140kVp result. On the other hand, the EF-Net shows a consistently good performance.

\begin{figure}
    \centering
    \includegraphics[scale=.13]{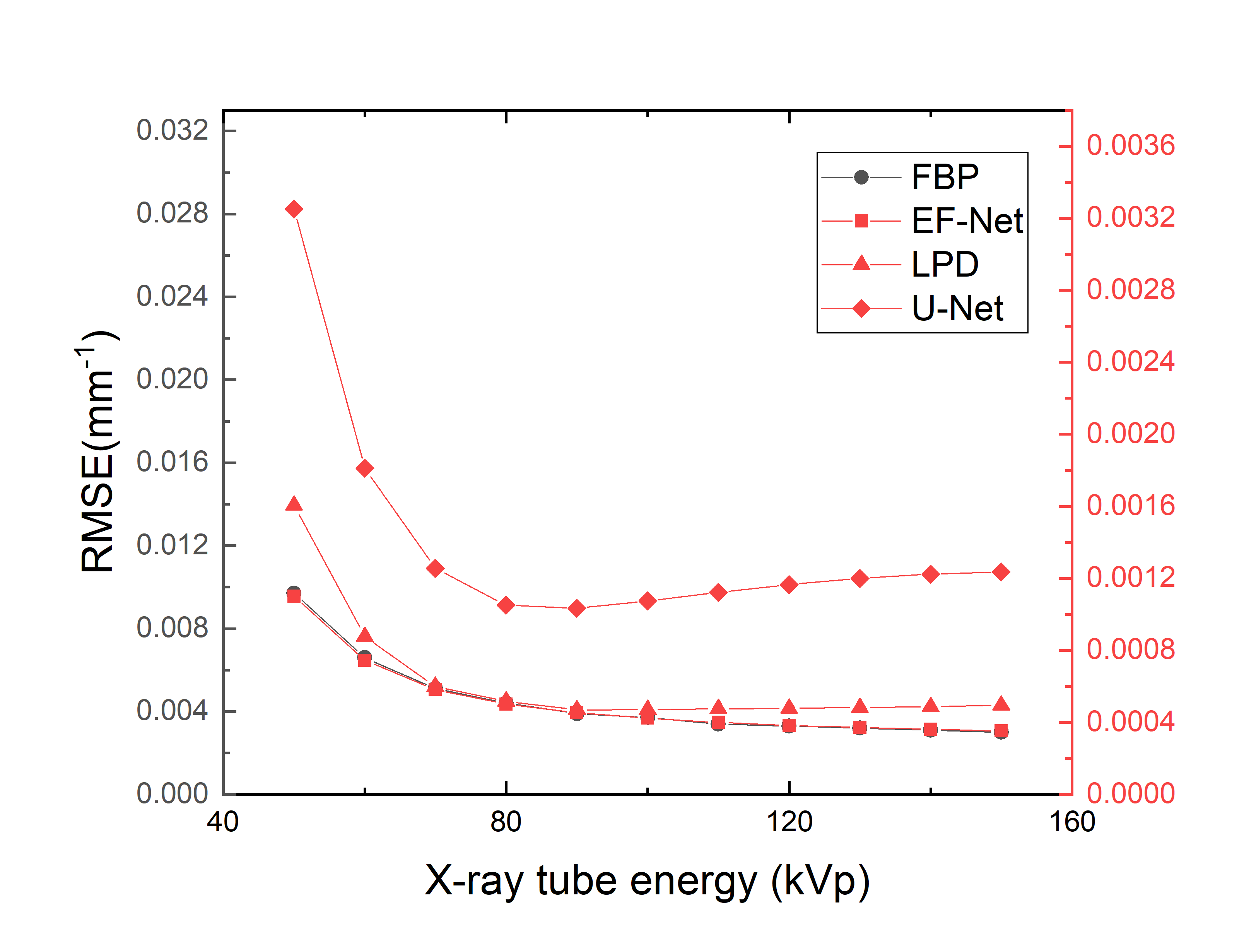}
    \caption{RMSEs of the reconstruction methods are plotted. The RMSEs were calculated for every 10kVp from 50 to 150kVp. The RMSEs of the FBP are plotted according to the left Y-axis, and the others are plotted according to the right Y-axis.}
    \label{fig10}
\end{figure}

Furthermore, we investigated the energy flexibility of the EF-Net for a broad X-ray energy range: 50 to 150kVp.  In Figure \ref{fig10}, we present the RMSEs at various X-ray tube voltages for the chest CT data. As in the previous study, all networks except the EF-Net were trained by 80kVp training data. Please note that the RMSEs of the FBP results use the scale on the left axis and other ones use the right axis scale. While the FBP has nearly 10 times higher RMSEs than the EF-Net, the trends of RMSEs of the FBP and the EF-Net are very similar within the trained energy range, 80 to 140kVp. Surprisingly, the similarity is fairly preserved even at the exterior energies to the trained energy range. On the contrary, the trends of the RMSEs of the U-Net and the LPD are significantly different from the FBP and the EF-Net, and the RMSEs of them become gradually worse as the test energy is away from the trained energy, 80kVp.

\section{Discussion}
In the above results, the performances of the FBP and the TV-GMI are not visually affected by the change in X-ray tube voltage. However, the performances of the U-Net and the LPD are significantly degraded at the untrained X-ray tube voltage, 100 kVp. The farther the test tube voltage is set away from the trained energy, the worse the performance becomes. Comparing the two, the LPD showed not only higher performance at the trained energy but also less performance degradation than the U-Net at the untrained energy data. It is thought to be because the LPD has much fewer parameters than the U-Net, implying that the LPD less overfits to the tube voltage of the trained data than the U-Net.

On the other hand, the performance of the EF-Net is robust against changes in X-ray tube voltage within the trained energy range from 80 to 140kVp. It appears that the quantitative results of the 100kVp test data are slightly worse than those of the 140kVp test data for the EF-Net in both the XCAT data and the chest data. However, it is due to the scale difference on bone attenuation coefficients between 100kVp and 140kVp and is not considered the performance degradation. This is also supported by Figure \ref{fig10} where the RMSEs for both the FBP and the EF-Net decrease as X-ray tube energy increases.

From the network structure point-of-view, the noticeable difference between the proposed network and the LPD can be found in the FAN and the second subnetwork while both networks have dual-domain structures and use almost the same number of parameters. By comparing the quantitative results of EF-Net without the EFT and LPD, we showed that recurring with the same parameters at every iteration and intensively regularizing the reconstructed image through the FAN and the second subnetwork is efficient to improve the network performance without significantly increasing the number of parameters.

This study was only conducted using the fan-beam simulation datasets. In clinical DECT data, however, it is difficult to separate the effect of X-ray tube voltage changes from other effects, such as X-ray scattering, anti-scatter grids, bowtie filtration, etc. As a preliminary study, we focused on verifying the feasibility of the EFT against changes on X-ray tube energy by excluding other influences. To test the proposed method for clinical DECT data, the aforementioned effects should be separately addressed or some modifications in the proposed network to treat local spectral differences must be needed, but it is beyond the scope of this study and would be a next step.

\section{Conclusions}
We proposed a new training method for the energy flexibility of DL networks and developed an effective dual-domain convolutional network. The feasibility of the proposed training method is successfully demonstrated by showing the energy flexibility of the proposed network within the X-ray energy range between 80kVp to 140kVp in the two simulation datasets. In contrast to other networks, the proposed network showed its robustness against the X-ray energy of the input data even for untrained energies. We expect that the proposed training method can greatly reduce the burden of data acquisition for the use of DL networks in DECT.

\section*{Acknowledgments}
This work was supported in part by the National Research Foundation of Korea under NRF-2020R1A2C2011959, in part by the institute of Civil Military Technology Cooperation funded by the Defense Acquisition Program Administration and Ministry of Trade, Industry and Energy of Korean government under Grant UM19207RD2, in part by the Korea Medical Device Development Fund under Grant 1711137888, and in part by the Ministry of Trade, Industry and Energy (MOTIE, Korea) under Grant 20014921.

%%Harvard
\section*{References}
\bibliographystyle{unsrt}
\bibliography{refs}

\end{document}